# Gear-based Metamaterials for Extraordinary Bandgap Tunability

Xin Fang[1#], Jihong Wen[1], Dianlong Yu[1], Peter Gumbsch[2,3], Huajian Gao[4]

[1]College of Intelligent Science and Technology, National University of Defense Technology, Changsha, China.

[2]Institute for Applied Materials, Karlsruhe Institute of Technology, Karlsruhe, Germany.

[3]Fraunhofer Institute for Mechanics of Materials IWM, Freiburg, Germany.

[4]Mechano-X Institute, Applied Mechanics Laboratory, Department of Engineering Mechanics, Tsinghua University, Beijing, China

#Email: xinfangdr@sina.com

**Abstract:** Metamaterials can be engineered with tunable bandgaps to adapt to dynamic and complex environments, particularly for controlling elastic waves and vibration. However, achieving wide-range, seamless, reversible, in-situ and robust tunability remains challenging and often impractical due to limitations in bandgap mechanisms and design principles. Here, we introduce gear-based metamaterials with unprecedented bandgap tunability. Our approach leverages Taiji planetary gear systems as variable-frequency local resonators, which allows the metamaterial to seamlessly modulate its bandgap's center frequency by 3~7 times (e.g. shifting from 250-430 Hz to 1400-2000 Hz), surpassing existing methods. Notably, this is achieved without pre-deformation or major changes to its static stiffness in the wave propagation direction, ensuring robust in-situ tunability and smooth control even under heavy static loads. This enables adaptable wave manipulation for versatile smart platforms.

Mechanical metamaterials can be designed to achieve extraordinary properties that surpass those found in nature, including high stiffness[1,2], high elastic energy density [3], nearly perfect wave absorption[4-6], high sound insulation[7-9] and topological wave transport[10-12]. Numerous dynamic properties for wave manipulation originate from bandgaps generated by local resonance[13,14], which can block elastic wave[15,16] and control vibration and noise[15,16]. Alike to the remarkable adaptivity of biological materials[19], attributes of metamaterials can be further enhanced through tunability of stiffness and shape[20-24]. Tunable bandgaps are highly desirable for adaptivity to complex environments and expanding the wave-control capabilities[25-28].

Bandgap frequencies generally scale with the system mass and stiffness as $\sqrt{stiffness/mass}$ (Supplementary Materials). The tunable range is defined as $X=f_{high}/f_{low}-1$, where $f_{high}$ and $f_{low}$ represent the center frequencies of the highest and lowest bandgaps, respectively. Tuning bandgaps by $X$ times necessitates synchronously adjusting the stiffness or mass of all metacells by $(X+1)^2$ times—a much greater range and often impractical. Moreover, meta-structures are often embedded in mechanical systems to control waves while supporting heavy static loads. Therefore, achieving robust tunability requests the controller to avoid inducing significant deformation in the wave propagation direction (strain $\varepsilon_{wave}$) or interfering with load bearing, i.e., requesting in-situ control[29] ($\varepsilon_{wave}=0$) and interference-free actuation.

This has been achieved in metamaterials based on piezoelectric[30,31], magnetorheological[32-35] and opto-responsive materials [36,37], but their tunable ranges remain limited to $X=0.06$~$0.5$ (FIG. 6). While using shape-memory materials[38-40] and buckling elements[41-44] can slightly extend the tunable range to $X=0.5$~$1$, this comes along with large deformation $\varepsilon_{wave}=0.1$~$0.5$ or non-smooth varying processes. Therefore, achieving metamaterial bandgaps with a broad tunable range, a smooth and reversible tuning process remains challenging.

Taking gears or gear clusters as metacells introduces a new paradigm for designing tunable metamaterials[45],

enabling excellent properties such as continuous modulation of stiffness[45], shape morphing [45-47] and self-organized granular aggregates [48,49], but tunable bandgaps have never been reported, requesting a new principle for realizing the desired tunability with compact metacell. Here we propose a strategy that breakthroughs the limitation via integrating periodic planetary gear systems as variable-frequency local resonators (FIG. 1).

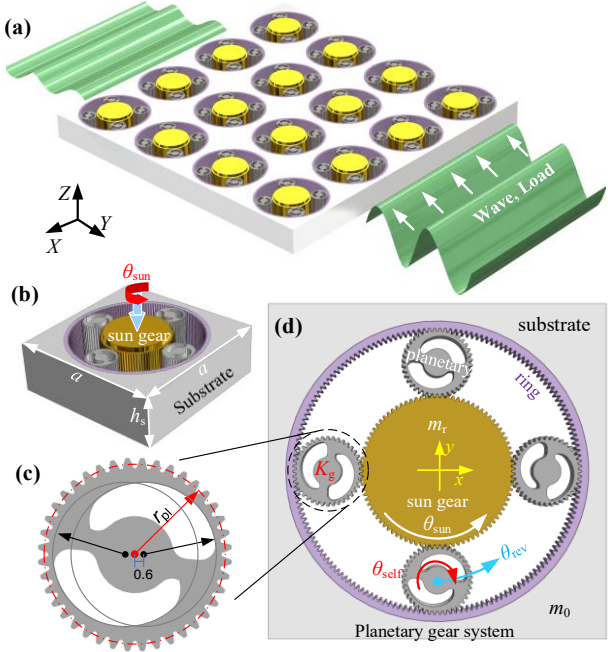

**FIG. 1. Configuration of gear-based bandgap metamaterials.** (a-b) Metamaterial structure. (c) Sectional view of the planetary gear. (d) Metacell structure, consisting of a substrate, a ring gear, a central sun gear, and four Taiji planetary gears. It also shows the revolution angle ($\theta_{rev}$) and self-rotation angle ($\theta_{self}$) of the planetary gears under the rotation actuation of sun gear ($\theta_{sun}$).

## Design Strategy

Different from existing gear-based metamaterial paradigm for tunable stiffness and shape[45], our bandgap metamaterial incorporates the Taiji planetary gear system as the compact variable-frequency local resonators, an principle that has not been explored previously (FIG. 1).

The local resonator depicted in FIG. 1b,d consists of a ring gear embedded within the substrate, a sun gear serving as the resonator mass $m_r$, and four symmetrical planetary gears that provide tunable stiffness $k_r$ for the resonator. To achieve a broad tunable range within a compact metacell, each planetary gear is designed with two center-symmetrical irregular apertures, forming two variable-thickness arms interconnected by a solid matrix

(FIG. 1c, Supplementary Materials). This configuration resembles a Taiji diagram, where the arms bend under radial compression. The bending stiffness of an individual arm $k_{arm}$ depends on the thickness at the meshing position, reaching a maximal $k_{arm}$ in the solid matrix and a minimal $k_{arm}$ near the thinnest section (FIG. 2). To ensure a synchronous stiffness modulation, the four planetary gears are precisely phase-aligned (FIG. 1d).

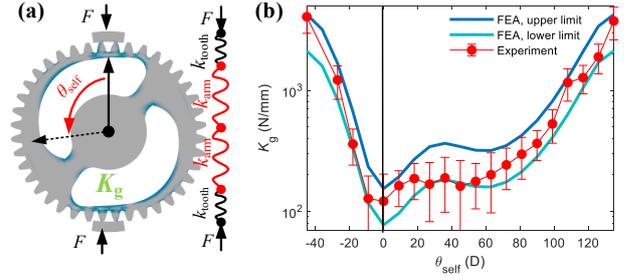

**FIG. 2 Variable stiffness of the Taiji planetary gear $K_g$.** (a) Mechanical models illustrating the composition of $K_g$. (b) Experimental and FEA results of $K_g$ under varying self-rotation angle $\theta_{self}$.

The radial stiffness of a planetary gear, *i.e.*, the connection stiffness between the ring and sun gears, $K_g=1/(2/k_{arm}+2/k_{tooth})$, is a function of $k_{arm}$ and meshing stiffness $k_{tooth}$ between gear teeth (FIG. 2). Although $k_{tooth}$ remains constant, when rotating the sun gear by angle $\theta_{sun}$, the planetary gears undergo self-rotation and revolution round the sun gear, changing the meshing position, modifying $k_{arm}$ and $K_g$, and ultimately tuning bandgap frequency. The revolution and self-rotation angles are

$$\theta_{rev} = \frac{\theta_{sun} r_{sun}}{2(r_{sun} + r_{pl})}, \quad \theta_{self} = -\frac{\theta_{sun} r_{sun}}{2 r_{pl}} \quad (1)$$

where $r_{sun}$ and $r_{pl}$ are pitch radii of the sun and planetary gears, respectively; the pitch radius of the ring gear is $r_{ring} = r_{sun}+2r_{pl}$. $\theta_{self} = 0$ is set at min($k_{arm}$). The origin $\theta_{sun} = 0$ is defined such that the planetary gears' center points align with the $x$ and $y$ axes at rest (FIG. 1d). Positive $\theta_{sun}$ denotes the right-handed rotation. The function $K_g(\theta_{self})$ is characterizable by the shape of Taiji diagram (FIG. 2b).

The ring is designed to be much stiffer than the substrate and the planetary gears, which can effectively transfer the dynamic force from the inner gears to the substrate while also providing a nearly constant static stiffness when tuning bandgaps. Finite element analyses (FEA) are performed to calculate the bands and wave modes of periodic metacell (FIG. 3a,b).

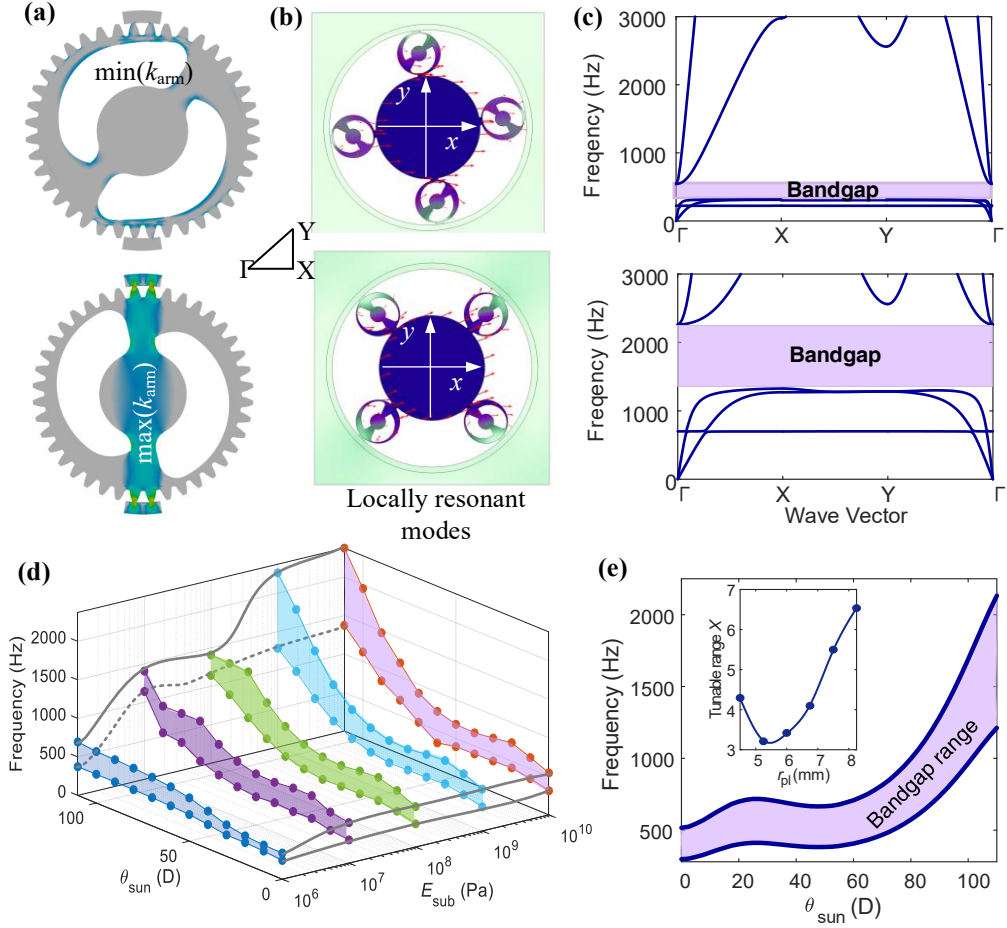

**FIG. 3. Tunable bandgap enabled by gear-based metamaterial.** (a) Deformation of a gear at $\theta_{self}=0°$ and 130°, corresponding to min($k_{arm}$) and max($k_{arm}$), respectively. (b) Stress fields showing the locally resonant modes for $\theta_{self}=0°$ and 130°, with little red arrows denote the vibration direction of the eigen modes. (c) Bands and bandgaps corresponding to (b). (d) Tunable bandgap by actuating the sun gear $\theta_{sun}$ for different substrate material modulus $E_{sub}$. (e) Tunable bandgap for $E_{sub}=1$ GPa picked from (d). A small panel is embedded into (e), showing the tunable range $X$ for different radius of planetary gears $r_{pl}$.

The bandgaps depend on the Young's modulus of the substrate material, $E_{sub}$, and the overall interaction stiffness between the sun and ring gears in the two-dimensional space. Here we focus on the first bandgap. For soft substrates, the first bandgap emerges due to Bragg scattering that is narrowly tunable (the blue ribbon in FIG. 3d, Supplementary Table 2). For moderate and stiff substrates, the first bandgap arises from locally resonant effect[13], in which the whole planetary system acts as a resonator inside the metacell (FIG. 3b). As $E_{sub}$ increases, the lowest bandgap at min($k_{arm}$) shifts to higher frequencies slowly, while the highest bandgap at max($k_{arm}$) rises rapidly, offering a broad range of tunability. For stiff substrate, reducing min($k_{arm}$) is the primary way for improving the tunable range $X$. This can be realized via reducing the thickness of the elastic arms. For specified minimal thickness, which is limited by structural integrity, increasing the radius of planetary gear $r_{pl}$ can also effectively expand the tunable range $X$ (FIG. 3e). For example, for specified ring's radius $r_{ring}=r_{sun}+2r_{pl}$, we can expand $X$ to 6.5 times by increasing planetary gear's radius $r_{pl}$ from 6 to 8.25 mm (FIG. 3e, Supplementary Materials).

**Metamaterial prototype**

To demonstrate the extraordinary tunability, we designed an unoptimized prototype consisting of square metacells with a lattice constant of $a=60$ mm (FIG. 1). The ring, planetary, and sun gears are fabricated from titanium, plastic, and copper, respectively (Supplementary

Materials), with tailored dimensions giving $\theta_{rev}=0.346\theta_{sun}$, $\theta_{self}=-1.125\theta_{sun}$ from Eq. (1). The minimal arm thickness of planetary gear is 0.3 mm.

We experimentally tested the variable stiffness $K_g$ of the plastic planetary gear (FIG. 2b). Due to contact nonlinearity in meshing, the upper and lower bounds of $K_g$ are defined by the elimination of clearance and the establishment of line contact between the teeth, respectively. The experimental results fall within the theoretical range. Notably, its stiffness can be modulated by 30 times through adjustments to the self-rotation angle $\theta_{self}$. Interestingly, despite a monotonically increasing arm thickness, $K_g$ exhibits a non-monotonic trend: It remains nearly constant within the range $20°<\theta_{self}<60°$ but increases rapidly within $60°<\theta_{self}<120°$.

The metamaterial shows tunable low-frequency bandgaps, with starting and cutoff frequencies ($f_{start}$, $f_{cut}$) modulated by substrate stiffness. For the soft substrate of $E_{sub}=1$ MPa, Bragg scattering governs the first bandgap, allowing for smooth but narrow modulation of bandgap's center frequency by $X=0.83$ times, from 244-339 Hz to 372-694 Hz (FIG. 3d). The normalized bandwidth of the bandgap, $w_{norm}=f_{cut}/f_{start}-1$, also increases from 0.39 to 0.77.

Locally resonant effect in the metamaterial with stiffer substrates offer broader tunable range, mainly attributed to the quickly improved cutoff frequency. When $E_{sub}>1$ GPa, further improving $E_{sub}$ can marginally improve the tunable range $X$ and the generalized bandwidth $w_{norm}$ because properties are dominated by the local resonance of gears. For the prototype with $E_{sub}=1$ GPa, we can smoothly shift bandgap from 313-543 Hz to 1283-2263 Hz, while the normalized bandwidth is constant $w_{norm}=0.75$. The varying trend of the bandgap frequency is completely consistent with the stiffness of the planetary gear $K_g$ (FIG. 2b). Notably, the tunable range of the bandgap's center frequency is $X=3.14$. The whole varying band 313-2263 Hz spans seven octaves, expanding the narrow bandgap for specified structures to broad band.

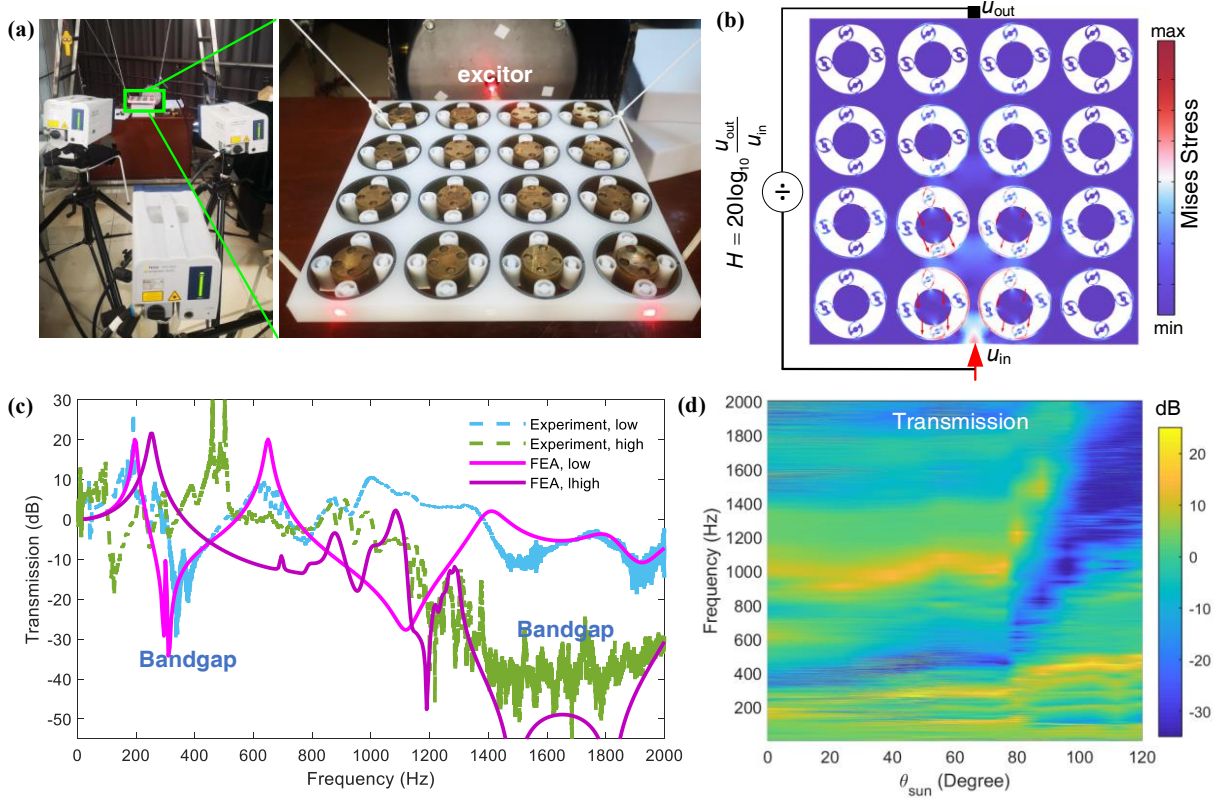

**FIG. 4. Experiments for broadly tunable bandgap.** (a) Experimental setup for testing the tunable bandgap metamaterial. (b) Typical wave field within the bandgap, illustrating the input and output signal locations. (c) Comparison of simulated and experimental transmission spectra for low- and high-frequency bandgaps. (d) Experimentally measured transmission spectra across varying rotation angle $\theta_{sun}$ of all sun gears. The blue region with great transmission loss indicates the bandgap range.

## Experiments

To demonstrate the capability of the gear-based metamaterial in achieving smooth, broad, and in-situ tunability of bandgaps, we constructed a metamaterial comprising a 4×4 array of square metacells (FIG. 4, Supplementary Materials). Longitudinal wave propagation within the metamaterial was measured using laser vibrometers. A sine-sweep signal $u_{in}$ was input at the center of one edge, and the corresponding response $u_{out}$ was measured at the center of the opposite edge (FIG. 4a,b). Wave transmission was quantified as $T=20\log_{10}(u_{out}/u_{in})$ dB, which varied as a function of the rotation angle $\theta_{sun}$ of all sun gears. FEA simulations of wave propagation were also conducted to validate the experimental results.

The experiment successfully captures the evolution of locally resonant bandgaps across varying angle $\theta_{sun}$ (FIG. 4d). The strong agreement between the experimental data and theoretical predictions highlights the robustness of our methodology (FIG. 4c). The bandgap frequency is smoothly tuned from 250-430 Hz to 1400-2000 Hz, with its center frequency shifting from 340 Hz to 1700 Hz, achieving a substantial tunability range of $X=4$.

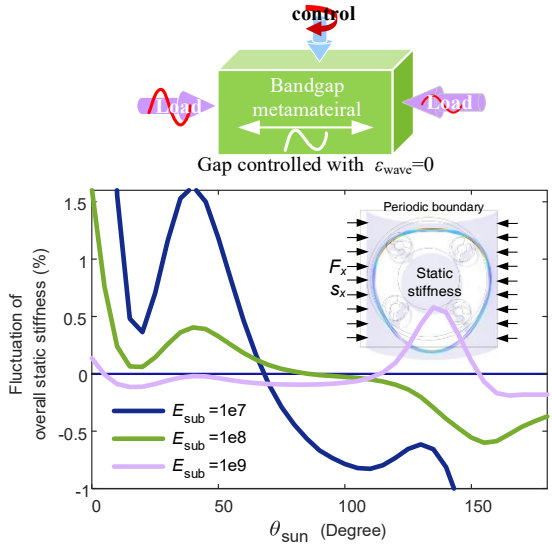

**FIG. 5 Fluctuation of the overall static stiffness of the metamaterial.** It illustrates that the bandgaps are controlled with $\varepsilon_{wave}=0$. In many conventional designs, bandgaps are controlled by large pre-deformation in the wave propagation direction ($\varepsilon_{wave}\gg 0$), causing interference between load bearing and bandgap control.

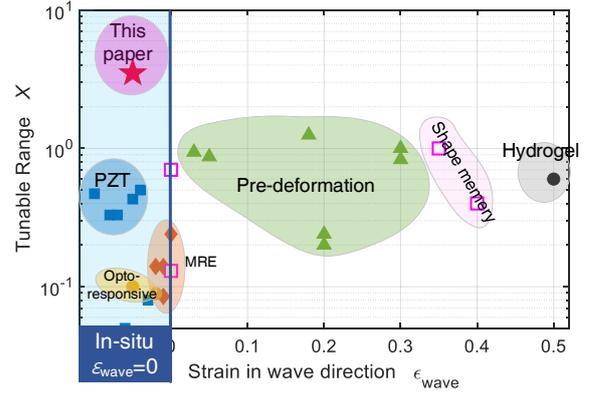

**FIG. 6. Performance of tunable bandgap metamaterials.** Comparison of tunable range $X$ versus the strain in the wave propagation direction $\varepsilon_{wave}$. Data includes various tunable metamaterial systems, such as piezoelectric (PZT), magnetorheological (MER), opto-responsive, hydrogel, shape-memory metamaterials, and the metamaterials consisting of buckling elements that request pre-deformation for tunability. Solid points represent smooth tunability, while empty points indicate non-smooth tunability. Details are explained in Supplementary Table 1.

## Discussions

As summarized in FIG. 6 and Supplementary Materials, existing tunable metamaterial bandgaps generally have narrow tunable ranges, or require large deformation in the wave propagation direction ($\varepsilon_{wave}\gg 0$) or exhibit non-smooth tuning processes.

To evaluate the static deformation of our new gear-based mechanical metamaterials, we calculate the metacell's static stiffness $K_{sx}$ in x-axis while applying continuity and periodic boundary conditions along the y-axis (FIG. 5). Then we evaluate the fluctuation of $K_{sx}$ using $[K_{sx}(\theta_{sun})/\text{average}(K_{sx})-1]\times 100\%$. A stiffer substrate offers narrower fluctuation range. For the substrate of $E_{sub}=1$ GPa, broad bandgap modulation (FIG. 3e) only marginally alters the overall static stiffness by 0.8% (FIG. 5). This means that we can smoothly modulate the dynamic properties with negligible alterations to static deformation in wave propagation direction under specified static loads, ensuring the in-situ tunability ($\varepsilon_{wave}=0$). This feature is essential for achieving seamless control under heavy static loads.

Additionally, the bandgap control is actuated out-of-plane rotation, effectively preventing interference between actuation and load-bearing functions. These advantages are critical for addressing the practical challenges of tunability in metamaterial design.

In summary, our approach allows the metamaterial to smoothly modulate the center frequency of its bandgap by more than $X=4$ times, significantly outperforming existing methods. The bandgap modulation avoids inducing deformation in the wave propagation direction ($\varepsilon_{wave}=0$), ensuring robust in-situ tunability, maintaining structural stability, interference-free control even under heavy static loads. By controlling the rotation angle of all sun gears synchronously or individually, our metamaterial can be utilized to develop tunable wave-insulation isolator[50], precise programmable wave manipulation[51,52] and mechanical computation[53,54], delivering advanced performance and robustness with wide-ranging potential applications.

## Acknowledgements
This research was funded by the National Natural Science Foundation of China (projects no. 52322505, and no. 52241103), the Natural Science Foundation of Hunan Province (projects no. 2023JJ10055).

## AUTHOR DECLARATIONS
**Conflict of Interest:**
A patent application has been submitted to China National Intellectual Property Administration. The patent has entered the substantive examination phase. The serial number is 2024110432791. The application date is 2024.07.31. Authors of this paper, Xin Fang, Jihong Wen and Dianlong Yu, are on the patent.

**Author contributions:**
Xin Fang: Conceptualization, Methodology, Investigation, Testing, Visualization, Supervision, Writing—original draft, Writing—review & editing. Jihong Wen: Methods, Supervision and Discussion. Dianlong Yu: Fabrication, testing and writing. Peter Gumbsch and Huajian Gao: Investigation, Supervision, Writing—original draft, Writing—review & editing.

**Data availability**
All data are available in the main text or the supplementary materials.

## REFERENCES
[1] X. Zheng et al., "Ultralight, ultrastiff mechanical metamaterials," Science **344**(6190)**,** 1373-1377 (2014).

[2] H. Guo and J. Zhang, "Performance-Oriented and Deformation-Constrained Dual-topology Metamaterial with High-Stress Uniformity and Extraordinary Plastic Property," Adv. Mater. **37**(7)**,** 2412064 (2025).

[3] X. Fang et al., "Large recoverable elastic energy in chiral metamaterials via twist buckling," Nature **639**(8055)**,** 639-645 (2025).

[4] T. Li et al., "Integrated adjustable acoustic metacage for multi-frequency noise reduction," Appl. Acoust. **217,** 109841 (2024).

[5] S. Qu and P. Sheng, "Microwave and Acoustic Absorption Metamaterials," Phys. Rev. Appl. **17**(4) (2022).

[6] X. Chen et al., "A compact acoustic metamaterial based on Helmholtz resonators with side slits for low-frequency sound absorption," Appl. Phys. Lett. **125**(1)**,** 13502 (2024).

[7] H. Zhang, S. Chen, Z. Liu, Y. Song, and Y. Xiao, "Light-weight large-scale tunable metamaterial panel for low-frequency sound insulation," Appl. Phys. Express **13**(6)**,** 67003 (2020).

[8] R. Ghaffarivardavagh, J. Nikolajczyk, S. Anderson, and X. Zhang, "Ultra-open acoustic metamaterial silencer based on Fano-like interference," Phys. Rev. B **99**(2)**,** 24302 (2019).

[9] H. Zhang, Y. Xiao, J. Wen, D. Yu, and X. Wen, "Ultra-thin smart acoustic metasurface for low-frequency sound insulation," Appl. Phys. Lett. **108**(14)**,** 141902 (2016).

[10] W. Wang, X. Wang, and G. Ma, "Non-Hermitian morphing of topological modes," Nature **608**(7921)**,** 50-55 (2022).

[11] X. Zhang, F. Zangeneh-Nejad, Z. G. Chen, M. H. Lu, and J. Christensen, "A second wave of topological phenomena in photonics and acoustics," Nature **618**(7966)**,** 687-697 (2023).

[12] C. Han et al., "Nonlocal Acoustic Moiré Hyperbolic Metasurfaces," Adv. Mater. **36**(18)**,** 2311350 (2024).

[13] Z. Liu et al., "Locally resonant sonic materials," Science **289**(5485), 1734-1736 (2000).

[14] S. Liu et al., "Experimental observation of negative rotational inertia," Appl. Phys. Lett. **123**(12), 121701 (2023).

[15] X. Fang, J. Wen, H. Benisty, and D. Yu, "Ultrabroad acoustical limiting in nonlinear metamaterials due to adaptive-broadening band-gap effect," Phys. Rev. B **101**(10), 104304 (2020).

[16] X. Fang, J. Wen, B. Bonello, J. Yin, and D. Yu, "Ultra-low and ultra-broad-band nonlinear acoustic metamaterials," Nat. Commun. **8**(1), 1288 (2017).

[17] G. Yan, Y. Li, Y. Wang, G. Yin, and S. Yao, "Tunable bandgap characteristic of various hexagon-type elastic metamaterials for broadband vibration attenuation," Aerosp. Sci. Technol. **145**, 108872 (2024).

[18] B. Hu, X. Fang, J. Wen, and D. Yu, "Effectively reduce transient vibration of 2D wing with bi-stable metamaterial," Int. J. Mech. Sci. **272**, 109172 (2024).

[19] Z. Wang et al., "Bio-inspired mechanically adaptive materials through vibration-induced crosslinking," Nat. Mater. **20**(6), 869-874 (2021).

[20] D. J. Levine, K. T. Turner, and J. H. Pikul, "Materials with Electroprogrammable Stiffness," Adv. Mater. **33**(35), 2007952 (2021).

[21] S. Leanza, S. Wu, X. Sun, H. J. Qi, and R. R. Zhao, "Active Materials for Functional Origami," Adv. Mater. **36**(9) (2024).

[22] X. Wang, Z. Meng, and C. Q. Chen, "Robotic Materials Transformable Between Elasticity and Plasticity," Adv. Sci. **10**(13) (2023).

[23] Y. Wang, Y. Wang, B. Wu, W. Chen, and Y. Wang, "Tunable and Active Phononic Crystals and Metamaterials," Appl. Mech. Rev. **72**(4), 40801 (2020).

[24] B. Wu et al., "Wave Manipulation in Intelligent Metamaterials: Recent Progress and Prospects," Adv. Funct. Mater. **34**(29), 2316745 (2024).

[25] S. M. Montgomery et al., "Magneto-Mechanical Metamaterials with Widely Tunable Mechanical Properties and Acoustic Bandgaps," Adv. Funct. Mater. **31**(3) (2021).

[26] X. Fei, L. Jin, X. Zhang, X. Li, and M. Lu, "Three-dimensional anti-chiral auxetic metamaterial with tunable phononic bandgap," Appl. Phys. Lett. **116**(2), 21902 (2020).

[27] S. Alan, A. Allam, and A. Erturk, "Programmable mode conversion and bandgap formation for surface acoustic waves using piezoelectric metamaterials," Appl. Phys. Lett. **115**(9), 93502 (2019).

[28] Y. Chen, B. Wu, M. Destrade, and W. Chen, "Voltage-controlled topological interface states for bending waves in soft dielectric phononic crystal plates," Int. J. Solids Struct. **259**, 112013 (2022).

[29] Z. Zhai, Y. Wang, K. Lin, L. Wu, and H. Jiang, "In situ stiffness manipulation using elegant curved origami," Sci. Adv. **6**(47), eabe2000 (2020).

[30] A. Bacigalupo, M. L. De Bellis, and D. Misseroni, "Design of tunable acoustic metamaterials with periodic piezoelectric microstructure," Extreme Mech. Lett. **40**, 100977 (2020).

[31] P. Zhang, H. Jia, Y. Yang, J. Wu, and J. Yang, "Extended topological interface modes with tunable frequency in the piezoelectric phononic crystal," Appl. Phys. Lett. **122**(18) (2023).

[32] Y. Xue, J. Li, Y. Wang, Z. Song, and A. O. Krushynska, "Widely tunable magnetorheological metamaterials with nonlinear amplification mechanism," Int. J. Mech. Sci. **264**, 108830 (2024).

[33] V. N. Gorshkov, O. V. Bereznykov, G. K. Boiger, P. Sareh, and A. S. Fallah, "Acoustic metamaterials with controllable bandgap gates based on magnetorheological elastomers," Int. J. Mech. Sci. **238**, 107829 (2023).

[34] Z. Chen et al., "Investigation of a new metamaterial magnetorheological elastomer isolator with tunable vibration bandgaps," Mech. Syst. Signal Proc. **170**, 108806 (2022).

[35] S. Liu, Y. Zhao, D. Zhao, J. Wu, and C. Gao, "Tunable Elastic wave Bandgaps and Waveguides by Acoustic Metamaterials with Magnetorheological Elastomer," Acoust. Phys. **66**(2), 123-131 (2020).

[36] A. S. Gliozzi et al., "Tunable photo-responsive elastic metamaterials," Nat. Commun. **11**(1), 2576 (2020).

[37] H. Patel, J. Chen, Y. Hu, and A. Erturk, "Photo-responsive hydrogel-based re-programmable metamaterials," Sci. Rep. **12**(1) (2022).

[38] X. Zhang *et al.*, "Inverse-designed flexural wave metamaterial beams with thermally induced tunability," Int. J. Mech. Sci. **267**, 109007 (2024).

[39] Y. Wei, Q. Yang, and R. Tao, "SMP-based chiral auxetic mechanical metamaterial with tunable bandgap function," Int. J. Mech. Sci. **195**, 106267 (2021).

[40] V. Candido De Sousa, D. Tan, C. De Marqui, and A. Erturk, "Tunable metamaterial beam with shape memory alloy resonators: Theory and experiment," Appl. Phys. Lett. **113**(14), 143502 (2018).

[41] H. Han, V. Sorokin, L. Tang, and D. Cao, "Origami-based tunable mechanical memory metamaterial for vibration attenuation," Mech. Syst. Signal Proc. **188**, 110033 (2023).

[42] E. Liu, X. Fang, P. Zhu, and J. Wen, "Stabilise and symmetrise the deformation of buckling metamaterial for tunable vibration bandgaps," Programmable Materials **1** (2023).

[43] K. Pajunen, P. Celli, and C. Daraio, "Prestrain-induced bandgap tuning in 3D-printed tensegrity-inspired lattice structures," Extreme Mech. Lett. **44**, 101236 (2021).

[44] M. Miniaci, M. Mazzotti, A. Amendola, and F. Fraternali, "Effect of prestress on phononic band gaps induced by inertial amplification," Int. J. Solids Struct. **216**, 156-166 (2021).

[45] X. Fang *et al.*, "Programmable gear-based mechanical metamaterials," Nat. Mater. **21**(8), 869-876 (2022).

[46] A. S. Meeussen, J. Paulose, and V. Vitelli, "Geared Topological Metamaterials with Tunable Mechanical Stability," Phys. Rev. X. **6**(4), 41029 (2016).

[47] F. Ma *et al.*, "Nonlinear Topological Mechanics in Elliptically Geared Isostatic Metamaterials," Phys. Rev. Lett. **131**(4) (2023).

[48] B. Saintyves, M. Spenko, and H. M. Jaeger, "A self-organizing robotic aggregate using solid and liquid-like collective states," Sci. Robot. **9**(86) (2024).

[49] A. Aubret, M. Youssef, S. Sacanna, and J. Palacci, "Targeted assembly and synchronization of self-spinning microgears," Nat. Phys. **14**(11), 1114-1118 (2018).

[50] W. Wei, F. Guan, and X. Fang, "A low-frequency and broadband wave-insulating vibration isolator based on plate-shaped metastructures," Appl. Math. Mech.-Engl. Ed. **45**(7), 1171-1188 (2024).

[51] S. Yu, C. Qiu, Y. Chong, S. Torquato, and N. Park, "Engineered disorder in photonics," Nat. Rev. Mater. **6**(3), 226-243 (2021).

[52] C. Cho, X. Wen, N. Park, and J. Li, "Digitally virtualized atoms for acoustic metamaterials," Nat. Commun. **11**(1), 251 (2020).

[53] M. Mousa and M. Nouh, "Parallel mechanical computing: Metamaterials that can multitask," Proc. Natl. Acad. Sci. U. S. A. **121**(52), e2407431121 (2024).

[54] N. Mohammadi Estakhri, B. Edwards, and N. Engheta, "Inverse-designed metastructures that solve equations," Science **363**(6433), 1333-1338 (2019).